\documentclass[12pt,a4paper]{article}
\usepackage[dvipdfm]{graphicx,color}
\usepackage{bm}
\usepackage{amsmath,amssymb}
\usepackage{booktabs}
\makeatletter
\def\Slash{\mathpalette\@Slash}
\def\@Slash#1#2{{\ooalign{\hfil$#1/$\hfil\crcr$#1{#2}$}}}
\makeatother
\begin{document}

\title{Nucleon decay via dimension-6 operators\\
in $E_6 \times SU(2)_F \times U(1)_A$ SUSY GUT model}
\author{
\centerline{
Nobuhiro~Maekawa$^{1,2}$\footnote{E-mail address: maekawa@eken.phys.nagoya-u.ac.jp}
~and 
Yu~Muramatsu$^{1}$\footnote{E-mail address: mura@eken.phys.nagoya-u.ac.jp}}
\\*[25pt]
\centerline{
\begin{minipage}{\linewidth}
\begin{center}
$^1${\it \normalsize Department of Physics, Nagoya University, Nagoya 464-8602, Japan }  \\*[10pt]
$^2${\it \normalsize Kobayashi Maskawa Institute, Nagoya University, Nagoya 464-8602, Japan }  \\*[10pt]\end{center}
\end{minipage}}
\\*[50pt]}
\date{}
\maketitle
\begin{abstract}
In the previous paper \cite{nucleon decay in U(1)A}, we have shown that 
$R_1\equiv\frac{\Gamma_{n \rightarrow \pi^0 + \nu^c}}{\Gamma_{p \rightarrow \pi^0 + e^c}}$ and 
$R_2\equiv\frac{\Gamma_{p \rightarrow K^0 + \mu^c}}{\Gamma_{p \rightarrow \pi^0 + e^c}}$ can
identify the grand unification group $SU(5)$, $SO(10)$, or $E_6$ in typical anomalous $U(1)_A$
supersymmetric (SUSY) grand unified theory (GUT) in which nucleon decay via dimension-6
operators becomes dominant. 
When $R_1 > 0.4$ the grand unification group is not $SU(5)$, while when 
$R_1 > 1$ the grand unification group is $E_6$.
Moreover, when $R_2 > 0.3$,  $E_6$ is implied.
Main ambiguities come from the diagonalizing matrices for quark and lepton
mass matrices in this calculation once we fix the vacuum expectation values of GUT Higgs bosons. 
In this paper, we calculate $R_1$ and $R_2$ in $E_6\times SU(2)_F$
SUSY GUT with anomalous $U(1)_A$ gauge symmetry, in which realistic quark and lepton masses and mixings
can be obtained though the flavor symmetry $SU(2)_F$ constrains Yukawa couplings at the GUT scale.
The ambiguities of Yukawa couplings are expected to be reduced.
We show that the predicted region for $R_1$ and $R_2$ is more restricted than in the $E_6$ model without
$SU(2)_F$ as expected.
Moreover, we re-examine the previous claim for the identification of grand unification group 
with $100$ times more model points ($10^6$ model points), including $E_6 \times SU(2)_F$ model.

\end{abstract}

\section{Introduction}
Grand unified theory (GUT) \cite{GUT} is one of the most favorable candidates for the model beyond the 
standard model (SM).
It has advantages not only theoretically but also experimentally.
Theoretical advantages are that it can unify the three gauge interactions in the SM 
into a single gauge interaction and particles in the SM into fewer multiplets.
Experimental advantages are that measured values of the three gauge couplings agree with 
the predicted values in supersymmetric (SUSY) GUT and measured hierarchies of masses 
and mixings of quarks and leptons can be understood, if it is assumed that ${\bf 10}$ matter induces 
stronger hierarchies for Yukawa couplings than the ${\bf \bar{5}}$ matter \cite{anarchy}.

The nucleon decay \cite{GUT,nucleon decay dim6,nucleon decay dim5} is one of the most important 
predictions in GUTs.
In GUTs there are new colored and $SU(2)_L$ doublet gauge bosons, which we call X-type gauge bosons.
In $SU(5)$ GUT models these gauge bosons are $X({\bf \bar{3}},{\bf 2})_{\frac{5}{6}}$ 
and $\bar{X}({\bf 3},{\bf 2})_{-\frac{5}{6}}$, where ${\bf \bar{3}}$ and ${\bf 2}$ 
means the antifundamental representation of $SU(3)_C$ and the fundamental representation of 
$SU(2)_L$, respectively, and $\frac{5}{6}$ means the hypercharge.
Exchanges of the X-type gauge bosons induce dimension-6 operators which break both the baryon 
and lepton numbers and induce the nucleon decay.
Usually, the main decay mode of the proton via dimension-6 operators is 
$p \rightarrow \pi^0 + e^c$.
The mass of the $X$ is roughly equal to the GUT scale at which three 
gauge couplings in the SM are unified into a single gauge coupling $g_{GUT}$, and 
therefore the lifetime of the nucleon can be estimated.
In the minimal SUSY GUT model, the GUT scale $\Lambda_G$ is $2 \times 10^{16}$ GeV, 
therefore the lifetime can be estimated as roughly $10^{36}$ years, which is
much larger than the current experimental lower bound, $10^{34}$ years \cite{SuperK}.

Triplet (colored) Higgs which is the GUT partner of the SM doublet Higgs also induces nucleon 
decay.
Because of smallness of Yukawa coupling for the first- and second-generation matters, 
the constraint on the triplet Higgs mass from the experimental limits 
of the nucleon lifetimes is not so severe without SUSY.
However, once SUSY is introduced, this constraint become severe because this induces 
nucleon decay via dimension-5 operators \cite{nucleon decay dim5}.
In the minimal $SU(5)$ SUSY GUT model, the lower bound for the triplet Higgs mass becomes 
larger than the GUT scale $\Lambda_G$ \cite{dim5bound,dim5bound'}.

The constraint on the triplet Higgs mass gives one of the most difficult problems 
in SUSY GUTs, i.e., the doublet-triplet splitting problem.
The SM doublet Higgs mass must be around the weak scale to realize electro-weak symmetry breaking, 
while as noted above, the GUT partner of that, 
triplet Higgs  must be heavier  than the GUT scale.
Of course, we can realize such a large mass splitting by fine-tunings, however it is 
unnatural.
A lot of attempts have been proposed to solve this problem \cite{DTsplitting review}.
However, in most of the solutions, some terms which are allowed by the symmetry are 
just neglected, or the coefficients for some terms are taken to be very small.
Such requirements are, in a sense, fine-tuning, and therefore, some mechanism which 
can realize such a large mass splitting in a natural way is required.

The doublet-triplet splitting problem can be solved under natural assumption 
by introducing anomalous $U(1)_A$ gauge symmetry.
The natural assumption means that all interactions which are allowed by symmetries of the models are 
introduced with $O(1)$ coefficients \cite{DTsplitting,GCUA,E6LLM,E6Higgs}.
Higher-dimensional interactions are also introduced if they are allowed by the symmetries.
One of the most interesting predictions of anomalous $U(1)_A$ SUSY GUT models is that
nucleon decay via dimension-6 operators becomes dominant \cite{GCUA}.
In these models the gauge coupling unification requires that the cutoff $\Lambda$ must be the usual 
SUSY GUT scale $\Lambda_G$ and the real GUT scale $\Lambda_u$ is
\begin{equation}
\Lambda_u \sim \lambda^{-a} \Lambda_G,
\end{equation}
where $\lambda < 1$ is the ratio of the Fayet-Iliopoulos (FI) parameter to cutoff $\Lambda$.
Because anomalous $U(1)_A$ charge of the adjoint Higgs $a$ is negative, $\Lambda_u$ is 
smaller than $\Lambda_G$, therefore, nucleon decay via dimension-6 operators is enhanced.
On the other hand, the nucleon decay via dimension-5 operators is strongly suppressed 
\cite{DTsplitting,GCUA}.
Therefore, the nucleon decay via dimension-6 effective operators is important in this scenario.
One more important feature is that 
the realistic quark and lepton masses and mixings can be realized in anomalous $U(1)_A$ 
SUSY GUT models, with $SO(10)$ and $E_6$ grand unification group \cite{DTsplitting,E6LLM}.

In the previous paper \cite{nucleon decay in U(1)A}, 
we have calculated various partial decay widths of nucleon from the
effective dimension-6 interactions in the anomalous $U(1)_A$ SUSY GUTs with $SU(5)$, $SO(10)$, or
$E_6$ grand unification group.
The predicted lifetime becomes just around the experimental lower bound, though the lifetime is strongly 
dependent on the explicit GUT models and the parameters. Therefore, it can happen that 
the nucleon decay is detected soon. The nucleon decay can be a good target for the future project.
It is difficult to kill the anomalous $U(1)_A$ GUT models from the limit of the lifetime
because the lifetime is proportional to the unification scale to the forth. 
However, we have claimed that the identification of the unification group in the anomalous $U(1)_A$
GUT scenario is possible if the several partial decay widths can be measured.
The ratio $R_1\equiv\frac{\Gamma_{n \rightarrow \pi^0 + \nu^c}}{\Gamma_{p \rightarrow \pi^0 + e^c}}$
is useful to know the largeness of the rank of the unification group because the contribution 
from the new X-type gauge bosons $X'$ in $SO(10)$ and $X''$ in $E_6$ make $R_1$ larger 
generically \cite{SU5 VS SO10}. 
And the ratio $R_2\equiv\frac{\Gamma_{p \rightarrow K^0 + \mu^c}}{\Gamma_{p \rightarrow \pi^0 + e^c}}$
is useful to catch the contribution from $X''$, 
which are mainly coupled  with the second generation fields of $\bf\bar 5$.
Note that these ratios are not dependent on the absolute values of 
vacuum expectation values (VEVs) of GUT Higgs bosons.
However, the results
are strongly dependent on the mass ratios of X-type gauge bosons.
It is important that the contribution from the extra gauge multiplet $X'$  becomes
always sizable in anomalous $U(1)_A$ GUT because the mass of $X'$ becomes 
almost the same as the mass of the $SU(5)$ superheavy gauge multiplet $X$. The contribution
from $X''$ can be large, though it is dependent on the explicit models. 
As a result, the identification becomes possible by measuring the ratios $R_1$ and $R_2$.
Once the masses of $X$ type gauge multiplets are fixed, the main ambiguities come from the diagonalizing
matrices of Yukawa matrices. These ambiguities cannot be fixed only from measured masses and mixings of quarks and leptons because we have a lot of $O(1)$ coefficients in the anomalous $U(1)_A$ GUT. 
If we would like to predict more concrete values for various decay modes of nucleons, we must fix
these $O(1)$ coefficients.

If we introduce the family symmetry $SU(2)_F$ into the anomalous $U(1)_A$ GUT with $E_6$ unification
group, the model predicts a characteristic scalar fermion mass spectrum in which 
the third generation $\bf 10_3$ of $SU(5)$ can have different universal sfermion masses $m_3$
from the other sfermions
which have universal sfermion masses $m_0$ \cite{E6SU2_1}. 
If we take $m_0>>m_3$, the SUSY 
flavor changing neutral current (FCNC) problem can be 
improved without destabilizing the weak scale because the FCNC constraints are weakened 
for large first two generation sfermion masses $m_0$ while the stop masses $m_3$, 
which is important for stabilization of the weak scale, can be around the weak scale.
In addition, if the CP symmetry is imposed, which is spontaneously broken by the Higgs which breaks
$SU(2)_F$, not only the SUSY CP problem can be solved but also the number of $O(1)$ coefficients for
quark and charged lepton masses and quark mixings can be smaller than the number of these mass and mixing 
parameters \cite{E6SU2_1_1,E6SU2_2,E6SU2_3}. It means that the diagonalizing matrices can be fixed from the quark and lepton
masses and mixings at least at the GUT scale in principle. In Ref. \cite{E6SU2_1_1}, 
it has been shown that the quark and charged lepton masses and the Cabibbo-Kobayashi-Maskawa (CKM) matrix
\cite{CKM} can be consistent with the values evaluated at the GUT scale in the minimal SUSY 
SM (MSSM) \cite{Ross} within factor 3 by choosing these parameters.
Once we could find the parameter set
at the GUT scale which realizes observed quark and lepton masses and mixings at the low energy scale 
in an explicit model, then we can predict various partial decay widths of the nucleon.

In this paper, we calculate the various decay widths of the nucleons in the 
$E_6\times SU(2)_F\times U(1)_A$ SUSY GUTs. 
If the parameter sets, which realize the observed quark and lepton masses and mixings, 
have been found easily, we would calculate the various decay widths by the parameter
sets.
However, it is not an easy task to find the parameter sets in calculating renormalization group 
equations (RGEs) which are dependent on the explicit GUT models. Alternatively, we find the relations 
between diagonalizing matrices which are independent of the renormalization scale, and under the 
relations we calculate the various decay widths of nucleon. Moreover, we re-examine the conditions
for the identification of the grand unification group by using 100 times more model points than in the
previous paper.

\section{$E_6 \times SU(2)_F \times U(1)_A$ SUSY GUT model}
\label{sec:def sec}
In this section we introduce the 
$E_6 \times SU(2)_F \times U(1)_A$ SUSY GUT model  \cite{E6SU2_1,E6SU2_1_1,E6SU2_2,E6SU2_3} and 
the diagonalizing matrices in the model are derived.
Setting and notation for the model in this paper are basically the same as these 
for the model in Ref. \cite{E6SU2_3}.
The diagonalizing matrix of light neutrinos, $L_\nu$, is derived from the Maki-Nakagawa-Sakata (MNS)
 matrix \cite{MNS} and the diagonalizing matrix of charged leptons, $L_e$, through the relation 
on the MNS matrix as $U_{MNS}=L_{\nu}^{\dagger} L_e$. 
Therefore we omit the explanation for the derivation of neutrino mass matrices in the model.
It is shown in Ref. \cite{E6SU2_3} in detail.

One of the most important features of the anomalous $U(1)_A$ gauge theory is that the VEVs 
of the GUT singlet operators $O_i$ are determined by their $U(1)_A$ charges 
$o_i$ as
\begin{equation}
\langle O_i\rangle=\left\{\begin{array}{c}  0   \qquad (o_i > 0) \\
                                           \lambda^{-o_i} \qquad (o_i \leq 0)
                         \end{array}\right.,
\end{equation} 
where $\lambda$ is determined from the FI parameter $\xi$ as $\lambda\equiv \xi/\Lambda$.
In this paper, we take $\lambda\sim 0.22$. As a result, the coefficient of the term $XYZ$ is
determined by their $U(1)_A$ charges, $x$, $y$, and $z$ as $\lambda^{x+y+z}XYZ$ if $x+y+z\geq 0$,
and they vanish if $x+y+z<0$. These features are important in understanding the following arguments
in this paper.

Contents of matters and Higgs and their charge assignment are shown in Table \ref{tb:Field contents}.
\begin{table}[tbp]
\begin{center}
\begin{tabular}{c|ccccccccccc}
 & $\Psi_a$ & $\Psi_3$ & $F_a$ & $\bar{F}^a$ & $\Phi$ & $\bar{\Phi}$ & $C$ & $\bar{C}$ & $A$ & $Z_3$ &$\Theta$\\
\hline
$E_6$ & ${\bf 27}$ & ${\bf 27}$ & ${\bf 1}$ & ${\bf 1}$ & ${\bf 27}$ & ${\bf \overline{27}}$ &
${\bf 27}$ & ${\bf \overline{27}}$ & ${\bf 78}$ & {\bf 1} & {\bf 1} \\
$SU(2)_F$ & ${\bf 2}$ & ${\bf 1}$ & ${\bf 2}$ & ${\bf \bar{2}}$ & ${\bf 1}$ & ${\bf 1}$ & 
${\bf 1}$ & ${\bf 1}$ & {\bf 1} & {\bf 1} &{\bf 1} \\
$U(1)_A$ & 4 & $\frac{3}{2}$ & -$\frac{3}{2}$ & -$\frac{5}{2}$ & -3 & 1 & -4 & -1 & -$\frac{1}{2}$ & -$\frac{3}{2}$ & -1\\
$Z_6$ & 3 & 3 & 1 & 0 & 0 & 0 & 5 & 0 & 0 & 0 & 0\\
\hline
\end{tabular}
\caption{Field contents and charge assignment under $E_6$$\times SU(2)_F\times U(1)_A\times Z_6$.}
\label{tb:Field contents}
\end{center}
\end{table}
In this paper, the capital letter denotes the superfield and the small letter
denotes the corresponding $U(1)_A$ charge.
$27$ dimensional (fundamental) representation of $E_6$ group, ${\bf 27}$, is decomposed in the 
$E_6\supset SO(10)\times U(1)_{V'}$ notation (and [$SO(10)\supset SU(5)\times U(1)_V$] 
notation) as 
\begin{equation}
\bm{27}=\bm{16}_1[\bm{10}_{1}+\bar{\bm{5}}_{-3}+\bm{1}_{5}]+
\bm{10}_{-2}[\bm{5}_{-2}+\bm{\bar{5}'}_{2}]+\bm{1}'_4[\bm{1}'_0].
\end{equation}
${\bf 16}$ and ${\bf 10}$ of $SO(10)$ are decomposed in the 
$SU(3)_C \times SU(2)_L \times U(1)_Y$ notation as
\begin{equation}
{\bf 16}\rightarrow \underbrace{q_{L}({\bf 3,2})_{\frac{1}{6}}
+u^c_R({\bf \bar{3},1})_{-\frac{2}{3}}
+e^c_R({\bf 1,1})_1}_{{\bf 10}}+
\underbrace{d^c_R({\bf \bar{3},1})_{\frac{1}{3}}
+l_L({\bf 1,2})_{-\frac{1}{2}}}_{{\bf \bar{5}}}+
\underbrace{\nu^c_R({\bf 1,1})_0}_{{\bf 1}},
\end{equation}
\begin{equation}
{\bf 10}\rightarrow \underbrace{D^c_R({\bf \bar{3},1})_{\frac{1}{3}}
+L_L({\bf 1,2})_{-\frac{1}{2}}}_{{\bf \bar{5'}}}+
\underbrace{\overline{D^c_R}({\bf 3,1})_{-\frac{1}{3}}
+\overline{L_L}({\bf 1,2})_{\frac{1}{2}}}_{{\bf 5}}.
\end{equation}
${\bf 27}$ includes two ${\bf \bar{5}}$s and two ${\bf 1}$s.
This feature plays an important role in realizing realistic quark and lepton masses and mixings 
in this model.
$F_a$ and $\bar{F}^a$ are Higgs which obtain VEVs 
$\langle F_a \rangle$ and $\langle \bar{F}^a \rangle$ as 
\begin{equation}
\langle F_a \rangle \sim \left(
\begin{array}{c}
0  \\
e^{i \rho}\lambda^{-(f+\bar{f})/2} \Lambda
\end{array}
\right),
\langle \bar{F}^a \rangle \sim \left(
\begin{array}{c}
0  \\
\lambda^{-(f+\bar{f})/2} \Lambda
\end{array}
\right),
\end{equation}
and break $SU(2)_F$.
Hereafter, we use unit in which the cutoff $\Lambda$ is taken as $\Lambda=1$.
$\Phi$ and $\bar{\Phi}$ are Higgs which obtain VEVs 
$\langle \Phi \rangle$ and $\langle \bar{\Phi} \rangle$ in SO(10) singlet direction as 
$\langle{ \bf 1'}_\Phi \rangle = \langle {\bf 1'}_{\bar{\Phi}} \rangle = v_{\phi} \sim \lambda^{-(\phi + \bar{\phi})/2}$ 
and break $E_6$ into $SO(10)$.
$C$ and $\bar{C}$ are Higgs which obtain VEVs 
$\langle C \rangle$ and $\langle \bar{C} \rangle$ in SU(5) singlet direction as 
$\langle {\bf 16}_C \rangle = \langle{ \bf\overline{16}}_{\bar{C}} \rangle = v_c \sim \lambda^{-(c + \bar{c})/2}$ 
and break $SO(10)$ into $SU(5)$.
$A$ is an adjoint Higgs which is decomposed in $SO(10)\times U(1)_{V'}$ notation as
$\bf 78\rightarrow 45_0+16_{-3}+\overline{16}_3+1_0$. $A$ obtains 
Dimopoulos-Wilczek type VEV $\langle A \rangle $ \cite{DW form} as 
\begin{equation}
\langle {\bf 45}_A \rangle=i \sigma_2 \times \left(
\begin{array}{ccccc}
x & & & &  \\
 & x & & &  \\
 & & x & &  \\
 & & & 0 &  \\
 & & & & 0 
\end{array}
\right)
\end{equation}
which is proportional to $B-L$ charge.
This VEV of adjoint Higgs plays an important role in realizing the doublet-triplet splitting.
Here, $\sigma_i$ $(i=1,2,3)$ are the Pauli matrices.

This model includes the MSSM doublet Higgs $H_u$ and $H_d$ as
\begin{equation}
H_u \subset \bm{5}_{\Phi},
\end{equation}
\begin{equation}
H_d \subset \bar{\bm{5}}'_{\Phi}+\beta_He^{i\delta}\lambda^{0.5}\bar{\bm{5}}_C
\label{eq:Hd}
\end{equation}
\cite{E6SU2_1_1},
where $\beta_H$ is a real $O(1)$ coefficient.  $\delta$ is a complex phase and depends on the models.
Please refer to the papers \cite{DTsplitting,GCUA,E6Higgs} to understand how to realize the doublet-triplet splitting 
in a natural way. 
Yukawa couplings are derived from the superpotential,
\begin{eqnarray}
W_Y&=&(a\Psi_3\Psi_3+b\Psi_3\bar F^a\Psi_a+c\bar F^a\Psi_a\bar F^b\Psi_b)\Phi+d(\Psi_a,\Phi,\bar \Phi,A,Z_3,\Theta) \nonumber \\
&&+f'\Theta \bar F^a\Psi_a\bar F^b\Psi_bC+g'\Theta\Psi_3\bar F^a\Psi_aC,
\end{eqnarray}
where $a$, $b$, $c$, $f'$, and $g'$ are $O(1)$ coefficients, and $d(\Psi_a,\Phi,\bar \Phi,A,Z_3,\Theta)$ 
is a gauge invariant function  of $\Psi_a$, $\Phi$, $\bar\Phi$, $A$, $Z_3$, and $\Theta$ and it 
contributes to $\Psi_1\Psi_2\Phi$.
Note that the operator $\epsilon^{ab}\Psi_a\Psi_b\Phi$ is not allowed because of asymmetric feature of
$\epsilon^{ab}$, where
$\epsilon^{ab}$ ($\epsilon^{12}=-\epsilon^{21}=1$) is 
antisymmetric tensor of $SU(2)_F$ group.
Therefore, the function $d$ includes, for example, 
$\Theta^2Z_3\epsilon^{ab}\Psi_a A\Psi_b\Phi$, $\bar\Phi Z_3\epsilon^{ab}\Psi_a A\Psi_b\Phi\Phi$, $\cdots$.
This function contains the following terms by developing the VEVs of $A$, $\Phi$, $\bar\Phi$, $Z_3$, and $\Theta$:
\begin{equation}
d(\Psi_a,\Phi,\bar \Phi,A,Z_3,\Theta) \rightarrow 
\frac{2}{3} d_5\lambda^5\epsilon^{ab}D^c_{R\Psi_a}\overline{D^c_R}_{\Psi_b}\bm{1}'_{\Phi}, 
\ \frac{1}{3}d_q\lambda^5\epsilon^{ab}q_{L\Psi_a}u^c_{R\Psi_b}(L)_{\Phi}, \nonumber
\end{equation}
\begin{equation}
\frac{1}{3}d_q\lambda^5\epsilon^{ab}q_{L\Psi_a}d^c_{R\Psi_b}(\bar{L})_{\Phi}, 
\ -d_l\lambda^5\epsilon^{ab}l_{L\Psi_a}e^c_{R\Psi_b}(\bar{L})_{\Phi}, 
\ h\lambda^5\epsilon^{ab}l_{L\Psi_a}\nu^c_{R\Psi_b}(L)_{\Phi},
\label{eq:d function}
\end{equation}
where $d_5$, $d_q$, $d_l$, and $h$ are real $O(1)$ coefficients.
Note that the coefficients of the first 4 terms in Eq. (\ref{eq:d function}) are proportional to
 $B-L$ charge.
The reason is as follows.
The above argument on asymmetric feature can be applied into the terms 
$\epsilon^{ab}{\bf 16}_a {\bf 16}_b {\bf 10}_{\Phi}$ and 
$\epsilon^{ab}{\bf 10}_a {\bf 10}_b {\bf 1}_{\Phi}$ of 
$SO(10)$.
To obtain non-zero terms, they must pick up the breaking of $SO(10)$, i.e., 
the adjoint Higgs VEV $\langle A \rangle$ which is proportional to the $B-L$ charge.

Up-type quark Yukawa matrix $Y_u$ is derived as
\begin{equation}
Y_u=\left(
\begin{array}{ccc}
0 & \frac{1}{3}d_q \lambda^5 & 0 \\
-\frac{1}{3}d_q \lambda^5 & c\lambda^4 & b\lambda^2  \\
0 & b\lambda^2 & a 
\end{array}
\right) \equiv \left(
\begin{array}{ccc}
0 & \frac{1}{3}y_{u12}\lambda^5 & 0  \\
-\frac{1}{3}y_{u12}\lambda^5 & y_{u22}\lambda^4 & y_{u23}\lambda^2  \\
0 & y_{u23}\lambda^2 & y_{u33} 
\end{array}
\right), \label{eq:Yu}
\end{equation}
where $y_{uij}$ ($i,j=1,2,3$) is a real $O(1)$ coefficient.
The factor $\frac{1}{3}$ in $(Y_u)_{12}$ and $(Y_u)_{21}$ plays 
an important role in obtaining small up quark mass.

Next, we derive down-type quark and charged lepton Yukawa matrices.
Note that three ${\bf 27}$ matters of $E_6$ include six ${\bf \bar{5}}$s of $SU(5)$.
Three of six ${\bf \bar{5}}$s become superheavy with three ${\bf 5}$s after developing the VEVs of
$\Phi$ and $C$.
The other three ${\bf \bar{5}}$s are massless, which are corresponding to the SM ${\bf \bar{5}}$s.
To obtain the SM ${\bf \bar{5}}$s, we estimate the mass matrix for ${\bf 5}$ and ${\bf \bar{5}}$.
Suppose the relation
\begin{equation}
\frac{\lambda^c\langle C\rangle}{\lambda^{\phi }\langle \Phi \rangle} 
=r \lambda^{0.5},
\end{equation}
which is important in obtaining the realistic large neutrino mixings.
Here $r$ is a real $O(1)$ coefficient.
Then, the mass matrix for ${\bf 5}$ and ${\bf \bar{5}}$ is derived as
\begin{equation}
\left(
\begin{array}{ccc|ccc}
0 & \alpha d_5\lambda^5 & 0 & 0 & fe^{i\rho}\lambda^{5.5} & ge^{i\rho}\lambda^{3.5}\\
-\alpha d_5\lambda^5 & c\lambda^4 & b\lambda^2 & fe^{i\rho}\lambda^{5.5} & 0 & 0\\
0 & b\lambda^2 & a & ge^{i\rho}\lambda^{3.5} & 0 & 0
\end{array}
\right)
\equiv  \left( \begin{array}{c|c}
M_1 & M_2
\end{array}
\right),
\label{eq:5_mass_matrix}
\end{equation}
where we re-define real $O(1)$ parameters $f'$ and $g'$ as $f \equiv r f'$ and $g \equiv r g'$.
Here, $\alpha=1$ for triplet (colored) component and $\alpha=0$ for doublet component.
We diagonalize the $3 \times 6$ mass matrix $(M_1\ M_2)$ as 
\begin{equation}
V^\dagger(M_1\ M_2)
\left(\begin{array}{cc}
U_{10}^H & U_{10}^0\\
U_{16}^H & U_{16}^0
\end{array}\right)=(M_H^{\rm{diag}}\ 0),
\end{equation}
where $V$ is $3 \times 3$ unitary matrix and $U$ is $6 \times 6$ unitary matrix which is given 
as
\begin{equation}
U\equiv\left(\begin{array}{cc}
U_{10}^H & U_{10}^0\\
U_{16}^H & U_{16}^0
\end{array}\right).
\end{equation}
The massless ${\bf \bar{5}_i^0}$ are given as
\begin{equation}
\bar{\bm{5}}^0_i\equiv(U_{10}^{0\dagger})_{ij}\bar{\bm{5}}'_j+(U_{16}^{0\dagger})_{ij}\bar{\bm{5}}_j
=\left(\begin{array}{c}
\bar{\bm{5}}_1+\cdots\\ \bar{\bm{5}}'_1+\cdots\\ \bar{\bm{5}}_2+\cdots
\end{array}\right),
\end{equation}
where $U_{10}^0$ and $U_{16}^0$ are calculated as
\begin{equation}
U_{10}^0=\left(\begin{array}{ccc}
-\frac{a\alpha d_5(bg-af)}{(ac-b^2)^2}\lambda^{2.5}e^{i\rho} & 1 & {\cal O}(\lambda^{5.5})\\
\frac{bg-af}{ac-b^2}\lambda^{1.5}e^{i\rho} & \frac{a\alpha d_5}{ac-b^2}\lambda &{\cal O}(\lambda^{4.5})\\
-(\frac{g}{a}+\frac{b}{a}\frac{bg-af}{ac-b^2})\lambda^{3.5}e^{i\rho} & -\frac{b\alpha d_5}{ac-b^2}\lambda^3 & {\cal O}(\lambda^{6.5})
\end{array}\right),
\end{equation}
\begin{equation}
U_{16}^0=\left(\begin{array}{ccc}
1 & 0 & 0\\
{\cal O}(\lambda^6) & 0 & 1\\
-\frac{bg-af}{ac-b^2}\frac{\alpha d_5}{g}\lambda^3 & -\frac{\alpha d_5^2}{ac-b^2}\frac{a}{g}\lambda^{2.5}e^{-i\rho} & -\frac{f}{g}\lambda^2
\end{array}\right).
\end{equation}
The detail derivation is shown in Ref. \cite{E6SU2_2, E6SU2_3}.

As a result, the down-type Yukawa matrix  $Y_d$ is given as
\begin{align}
Y_d &= Y^{\Phi} U_{16}^0 + \beta_H e^{-i\delta}\lambda^{0.5} Y^C U_{10}^0 \nonumber \\
 & =\left(
\begin{array}{c}
[\frac{bg-af}{ac-b^2}(f'-\frac{bg'}{a})-\frac{gg'}{a}]
\beta_H e^{i(2\rho-\delta)}\lambda^6    \\
\left( -\frac{d_q}{3}-\frac{bg-af}{ac-b^2} \frac{b \frac{2}{3}d_5}{g} \right) \lambda^5  \\
-\frac{bg-af}{ac-b^2} \frac{a \frac{2}{3}d_5}{g} \lambda^3 
\end{array} \right. \nonumber \\
& \hspace{1cm} \left.
\begin{array}{cc} 
-\frac{bg'-af'}{ac-b^2}\frac{2}{3}d_5 \beta_H e^{i(\rho-\delta)} \lambda^{5.5} & 
\frac{1}{3}d_q \lambda^5 \\ 
\left( -\frac{(\frac{2}{3} d_5)^2}{ac-b^2}\frac{ab}{g}e^{-i\rho} + 
f' \beta_H e^{i(\rho-\delta)} \right) \lambda^{4.5} & 
\left( \frac{ac-b^2}{a}+\frac{bg-af}{g}\frac{b}{a} \right) \lambda^4 \\
\left( -\frac{(\frac{2}{3} d_5)^2}{ac-b^2}\frac{a^2}{g}e^{-i\rho} + 
g' \beta_H e^{i(\rho-\delta)} \right) \lambda^{2.5} & 
\frac{bg-af}{g} \lambda^2
\end{array}\right) \nonumber \\
&\equiv \left(
\begin{array}{ccc}
y_{d11} \lambda^6 & \frac{2}{3}y_{d12} \lambda^{5.5} & \frac{1}{3} y_{d13} \lambda^5  \\
\frac{2}{3}y_{d21} \lambda^5 & y_{d22} \lambda^{4.5} & y_{d23} \lambda^4 \\
\frac{2}{3}y_{d31} \lambda^3 & y_{d32} \lambda^{2.5} & y_{d33} \lambda^2
\end{array}
\right), \label{eq:Yd}
\end{align}
where $y_{dij}$ is a $O(1)$ coefficient which includes complex phase.
In our calculation for nucleon decay $y_{dij}$ is taken to be a real $O(1)$ coefficient 
for simplicity.
Here,
\begin{equation}
Y^{\Phi}=\quad\left(
\begin{array}{ccc}
0 & -\frac{1}{2} d_q \lambda^5 & 0 \\
\frac{1}{2} d_q \lambda^5 & c\lambda^4 & b \lambda^2 \\
0 & b \lambda^2 & a
\end{array}\right),\;  
Y^C=\quad\left(
\begin{array}{ccc}
0 & f'e^{i\rho}\lambda^4 & g'e^{i\rho}\lambda^2 \\
f'e^{i\rho}\lambda^4 & 0 & 0 \\
g'e^{i\rho}\lambda^2 & 0 & 0
\end{array}\right).
\label{eq:YPhi}
\end{equation}

The charged lepton Yukawa matrix $Y_e$ is derived 
from a relationship $Y_e=Y_d^T$ with $\alpha=0$ ($d_5=0$) and $d_q/3 \rightarrow -d_l$
as
\begin{align}
Y_e& =\left(
\begin{array}{cc}
[\frac{bg-af}{ac-b^2}(f'-\frac{bg'}{a})-\frac{gg'}{a}] 
\beta_H e^{i(2\rho-\delta)} \lambda^6 & 
d_l \lambda^5 \\
0 & f' \beta_H e^{i(\rho-\delta)} \lambda^{4.5}  \\
-d_l \lambda^5 & \left( \frac{ac-b^2}{a} + 
\frac{bg-af}{g}\frac{b}{a} \right) \lambda^4 
\end{array}\right. \nonumber \\
& \hspace{8.3cm} \left.
\begin{array}{c}
0 \\
g' \beta_H e^{i(\rho-\delta)} \lambda^{2.5} \\
\frac{bg-af}{g} \lambda^2
\end{array}
\right) \nonumber \\
&\equiv \left(
\begin{array}{ccc}
y_{d11} \lambda^6 & y_{e12} \lambda^5 & 0  \\
0 & y_{e22} \lambda^{4.5} & y_{e23} \lambda^{2.5} \\
-y_{e12} \lambda^5 & y_{d23} \lambda^4 & y_{d33} \lambda^2
\end{array}
\right) \label{eq:Ye}
\end{align}
where $y_{eij}$ is a $O(1)$ coefficient which includes complex phase.
Again, we take $y_{eij}$ as a real $O(1)$ coefficient for simplicity.
Finally, to obtain $Y_u$, $Y_d$, and $Y_e$ we use 16 real parameters, $y_{uij}$, $y_{dij}$, and 
$y_{eij}$.
In the original $E_6\times SU(2)_F\times U(1)_A$ models, we have 9 real parameters and 2 CP phases.
Therefore, we have several relations among $y_{uij}$, $y_{dij}$, and $y_{eij}$.
We will discuss these relations in the next section.

Let us diagonalize these Yukawa matrices by field redefinition as
\begin{align}
\psi_{L i} Y_{ij} \psi_{Rj}^c &= 
(L_{\psi}^{\dagger} \psi_{L})_i (L_{\psi}^T Y R_{\psi})_{ij} (R_{\psi}^{\dagger}\psi_{R}^c)_j \\
 &=\psi'_{L i} Y_{diag\: ij} \psi'^c_{Rj}, \nonumber
\end{align}
where $\psi$ is a gauge eigenstate field and $\psi'$ is a mass eigenstate field.
We summarize the detail calculation in Appendix \ref{sec:diagonalizing}.
The diagonalizing matrices are calculated as
\begin{equation}
L_u \sim \left(
\begin{array}{ccc}
1 & \frac{1}{3} \lambda & 0 \\
\frac{1}{3} \lambda & 1 & \lambda^2  \\
\frac{1}{3} \lambda^3 & \lambda^2 & 1 
\end{array}
\right),R_u \sim \left(
\begin{array}{ccc}
1 & \frac{1}{3} \lambda & 0 \\
\frac{1}{3} \lambda & 1 & \lambda^2  \\
\frac{1}{3} \lambda^3 & \lambda^2 & 1 
\end{array}
\right),
\label{eq:diagonalizing up}
\end{equation}

\begin{equation}
L_d \sim \left(
\begin{array}{ccc}
1 & \frac{2}{3} \lambda & \frac{1}{3} \lambda^3 \\
\frac{2}{3} \lambda & 1 & \lambda^2  \\
\frac{2}{3} \lambda^3 & \lambda^2 & 1 
\end{array}
\right),R_d \sim \left(
\begin{array}{ccc}
1 & \frac{2}{3} \lambda^{0.5} & \frac{2}{3} \lambda \\
\frac{2}{3} \lambda^{0.5} & 1 & \lambda^{0.5}  \\
\frac{2}{3} \lambda & \lambda^{0.5} & 1 
\end{array}
\right),
\end{equation}

\begin{equation}
L_e \sim \left(
\begin{array}{ccc}
1 & \lambda^{0.5} & 0 \\
\lambda^{0.5} & 1 & \lambda^{0.5}  \\
\lambda & \lambda^{0.5} & 1 
\end{array}
\right),R_e \sim \left(
\begin{array}{ccc}
1 & \lambda & \lambda^3 \\
\lambda & 1 & \lambda^2  \\
\lambda^3 & \lambda^2 & 1 
\end{array}
\right),
\label{eq:diagonalizing electoron}
\end{equation}

\begin{equation}
L_{\nu} \sim \left(
\begin{array}{ccc}
1 & \lambda^{0.5} & \lambda \\
\lambda^{0.5} & 1 & \lambda^{0.5}  \\
\lambda & \lambda^{0.5} & 1 
\end{array}
\right).
\label{eq:diagonalizing neutrino}
\end{equation}
Since this model has a lot of $O(1)$ parameters for the right-handed neutrino mass matrix,
we do not have any interesting relations in $L_\nu$.
The realistic CKM and MNS matrices can be obtained as
\begin{equation}
U_{CKM }=L_u^{\dagger}L_d\sim \left(
\begin{array}{ccc}
1 & \frac{2}{3} \lambda & \lambda^4 \\
\frac{2}{3} \lambda & 1 & \lambda^2  \\
\frac{2}{3} \lambda^3 & \lambda^2 & 1 
\end{array}
\right),
U_{MNS}=L_{\nu}^{\dagger}L_e\sim \left(
\begin{array}{ccc}
1 & \lambda^{0.5} & \lambda \\
\lambda^{0.5} & 1 & \lambda^{0.5}  \\
\lambda & \lambda^{0.5} & 1 
\end{array}
\right),
\end{equation}
if we consider the $O(1)$ coefficients.
Since the coefficient of $(U_{CKM})_{13}$ is vanishing in leading order
in this model \cite{E6SU2_2}, the sub-leading contribution $\lambda^4$ is dominant.
As noted previously, we estimate $L_\nu$ from the observed $U_{MNS}$ and $L_e$.

\section{Conditions for the diagonalizing matrices}
\label{sec:renormalization}
In the original $E_6\times SU(2)_F\times U(1)_A$ SUSY GUT models with the spontaneously broken CP
 symmetry, the number of parameters for the Yukawa couplings of up quarks, down quarks, and charged
 leptons
is 9 (real parameters)+2 (CP phases), which is smaller than the number of observed parameters of masses 
and mixings.
Therefore, once we fix these parameters from the observed values of masses and mixings,
we can predict all diagonalizing matrices.
Main obstacle for this approach is that these Yukawa couplings are determined at the GUT scale.
If the masses and mixings at the GUT scale have been calculated from these measured parameters 
through the renormalization group equations (RGEs), we would adopt this approach.
Unfortunately, many new couplings, which can contribute the running of the Yukawa couplings, 
appear, when superheavy fields appear at the mass scales which are dependent on the models.
Of course, once we fix the GUT models, we can calculate the low energy effective theory. 
But it is not an easy task to fix the 11 parameters to satisfy the measured quark and lepton masses
and mixings by RGEs which change when superheavy fields decouple, though we can do it in principle. 

Therefore, in this paper, we adopt another approach. 
We select several relations between Yukawa couplings which are not strongly dependent on the
renormalization scale. Using these relations, we reduce the number of parameters.

As noted in the previous section, we consider real Yukawa couplings for simplicity.
Then, generically, we have 27 parameters for the Yukawa couplings of up quarks, down quarks,
and charged leptons. In the previous section, we introduced 16 real parameters $y_{uij}$, 
$y_{dij}$, and $y_{eij}$ for these Yukawa couplings. Therefore, there must be 11 $(=27-16)$ relations
among the parameters of masses and diagonalizing matrices. In the followings, the notation of the angles
are defined in the appendix.
From $(Y_u)_{13}=(Y_u)_{31}=(Y_e)_{13}=0$, 
the relations 
\begin{equation}
s_{13}^{uL}=0,\: s_{13}^{uR}=0,\: s_{13}^{eL}=0. \label{eq:con1}
\end{equation}
are derived, respectively. 
$(Y_u)_{23}=(Y_u)_{32}$, $(Y_u)_{12}=-(Y_u)_{21}$, and $(Y_u)_{11}=0$ result in
\begin{equation}
s_{23}^{uL}=s_{23}^{uR},\: s_{12}^{uL}=-s_{12}^{uR},\: (s_{12}^{uL})^2=m_u/m_c.
\label{eq:con2}
\end{equation}
$(Y_e)_{31}=-(Y_e)_{12}$ and $(Y_e)_{21}=0$ lead to 
\begin{equation}
s_{12}^{eR}=s_{23}^{eL}s_{12}^{eL},\: s_{13}^{eR}=-s_{12}^{eL}m_{\mu}/m_{\tau}.
\label{eq:con3}
\end{equation}
From the relations $(Y_d)_{33}=(Y_e)_{33}$ and $(Y_d)_{23}=(Y_e)_{32}$,
\begin{equation}
s_{23}^{dL}=s_{23}^{eR},\: m_b=m_\tau,
\label{eq:con4}
\end{equation}
are derived. The relation $m_b=m_\tau$ is useless for fixing the diagonalizing matrices.
Finally, when we think the relation $(Y_e)_{11}=(Y_d)_{11}$ in addition to the above relations,
we obtain
\begin{equation}
s_{13}^{eL}s_{13}^{eR}m_{\tau}+s_{12}^{eL}s_{12}^{eR}m_{\mu}+m_e=
s_{13}^{dL}s_{13}^{dR}m_{b}+s_{12}^{dL}s_{12}^{dR}m_{s}+m_d.
\label{eq:conf}
\end{equation}
In our analysis, we do not use the last relation because it is strongly dependent on the
renormalization scale. As a result, we use 9 relations in our analysis. 
We have checked the scale dependence of these relations by explicit numerical calculations
of the RGEs in the MSSM \cite{Yukawa RGeq MSSM,Yukawa RGeq SM}.

We have additional $7$ $(=16-9)$ relations because the original models have only 9 real parameters
($a$, $b$, $c$, $d_q$, $d_5$, $d_l$ $f$, $g$, and $\beta_H$) if we take vanishing CP phases. 
Unfortunately, these are strongly dependent on the renormalization scale, and therefore,
we do not use these relations in our analysis. 

As a result, we use only 6 parameters in our numerical calculations of nucleon decays
for 7 diagonalizing matrices $L_u$, $L_d$, $L_e$, $L_\nu$, $R_u$, $R_d$, $R_e$.
Since we assume real diagonalizing matrices, each matrix has three real parameters, generically.
The CKM matrix and the MNS matrix reduce the 21 parameters to 15 parameters, and because of 9 relations,
only $6$ $(=15-9)$ parameters are sufficient. (Strictly speaking, the signature of $s_{12}^{uL}$ is an 
additional parameter because the relation $(s_{12}^{uL})^2=m_u/m_c$ cannot fix the signature.)
Note that we have used 12 parameters for fixing the real diagonalizing matrices of the 
$E_6\times U(1)_A$ GUT models in the previous paper. 
In this paper, we have succeeded to reduce the number of parameters to half.

\section{Numerical calculation}
In this section, we calculate various partial decay widths of nucleons numerically.
And we compare the results with those
in the previous paper \cite{nucleon decay in U(1)A}.

In our calculation, we use the VEVs
\begin{equation}
x=1\times 10^{16}\:  \text{GeV},\quad v_c=5\times 10^{14}\: \text{GeV},\quad 
v_\phi=5\times 10^{15}\: \text{GeV},
\end{equation}
where $x$ is the scale of the adjoint Higgs VEV which breaks $E_6$ into 
$SU(3)_C\times SU(2)_L\times SU(2)_R\times U(1)_{B-L}\times U(1)_{V'}$,
$v_\phi$ is the VEV of $\Phi$ which breaks $U(1)_{V'}$, and $v_c$ is the Higgs VEV which
breaks $SU(2)_R\times U(1)_{B-L}$ into $U(1)_Y$.  
These VEVs are the same as VEVs of GUT Higgs adopted in the previous paper.
The larger $x$ leads to larger contribution of $SO(10)$ and $E_6$ superheavy gauge multiplets to
the nucleon decay processes. 
The X-type gauge boson masses are written as
\begin{equation}
M_X=g_{GUT}x,\quad M_{X'}=g_{GUT}\sqrt{x^2+v_c^2}, 
\quad M_{X''}=g_{GUT}\sqrt{\frac{x^2}{4}+v_{\phi}^2}.
\end{equation}

We generate the real diagonalizing matrices $L_u$, $L_d$, $L_e$, $L_\nu$, $R_u$, $R_d$, and $R_e$
as follows.
\begin{enumerate}
 \item Once the $\theta_{23}^{uL}$ $(s_{23}^{uL}=\sin \theta_{23}^{uL})$ is generated through the relation 
$\theta_{23}^{uL}=B_{23}^{uL}\lambda^2$ where $B_{23}^{uL}$ is the $O(1)$ coefficient determined 
randomly from 0.5 to 2, $L_u$ and $R_u$ can be fixed by the relations (\ref{eq:con1}) and (\ref{eq:con2}).
\item The three parameters for $R_d$ are generated randomly in the same manner as $\theta_{23}^{uL}$.
$L_d$ can be determined by $L_d=L_u U_{CKM}^{(exp)}$.
\item  Once we generate two parameters among six for $L_e$ and $R_e$ randomly, we can fix the other
4 parameters by the relations (\ref{eq:con1}), (\ref{eq:con3}), and (\ref{eq:con4}).
\item The $L_\nu$ can be determined by $L_{\nu}=L_e U_{MNS}^{(exp)}$.
\item We check whether all $O(1)$ coefficients $B_{ij}$ of the components of the diagonalizing matrices 
($L_u$, $L_d$, $L_e$, $L_\nu$ $R_u$, $R_d$, and  $R_e$) in 
 Eq. (\ref{eq:diagonalizing up})-(\ref{eq:diagonalizing neutrino}) are within the region 
 $0.5\leq B_{ij}\leq 2$ or not. We adopt the parameter set only if all coefficients satisfy the
 condition.
\end{enumerate}
In the above calculation we use $m_u/m_c=0.0021$, $m_\mu/m_\tau=0.059$ \cite{quark mass},  
\begin{equation}
{\footnotesize U_{CKM}^{(exp)}=\begin{pmatrix}
0.97 & 0.23 & 0.0035 \\
-0.23 & 0.97 & 0.041 \\
0.0086 & -0.040 & 1.0
\end{pmatrix},
U_{MNS}^{(exp)}=\begin{pmatrix}
0.83 & 0.54 & 0.15 \\
-0.48 & 0.53 & 0.70 \\
0.30 & -0.65 & 0.70
\end{pmatrix}},
\end{equation}
\cite{PDG,theta13}.
Following the above procedure, we have generated $10^4-10^6$ model points and calculated various 
partial decay widths of nucleons.

\subsection{Various decay modes for the proton}
\begin{figure}[htb]
 \centering
 \includegraphics[width=12cm,clip]{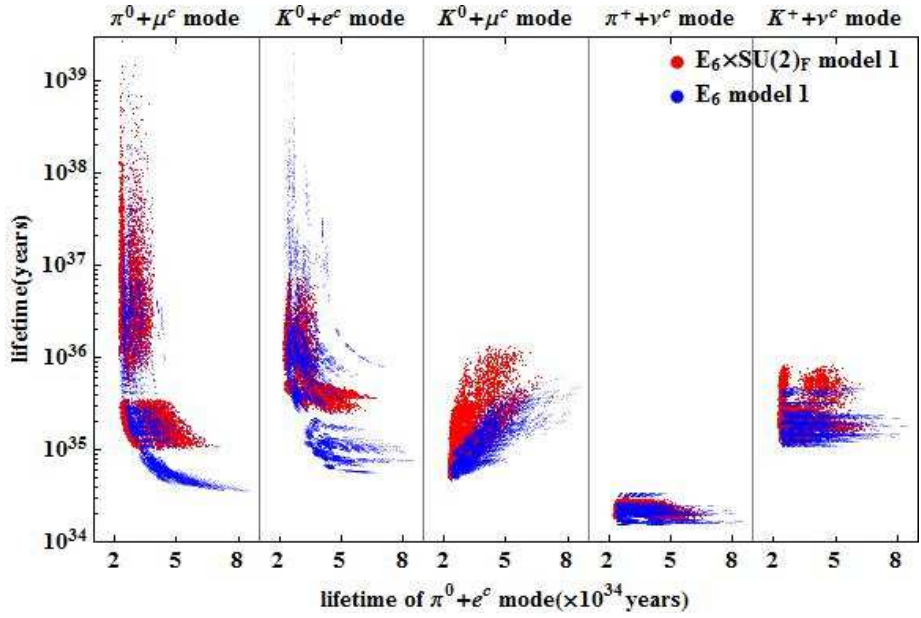}
 \caption{The proton lifetimes for various decay modes 
in the $E_6 \times SU(2)_F$ model 1 and the $E_6$ model 1 with 
$M_X=g_{GUT}x$, $M_{X'}=g_{GUT}\sqrt{x^2+v_c^2}$, 
$M_{X''}=g_{GUT}\sqrt{\frac{x^2}{4}+v_{\phi}^2}$, $x=1\times 10^{16}$ GeV, 
$v_c=5\times 10^{14}$ GeV, $v_{\phi}=5\times 10^{15}$ GeV.
Each model has $10^4$ model points.}
 \label{FigE6SU2}
\end{figure}

We calculate the proton lifetimes for various decay modes 
in the $E_6 \times SU(2)_F$ model.
As in the previous paper \cite{nucleon decay in U(1)A}, we use the hadron matrix elements
calculated by QCD lattice
\cite{form factor}.
The result are shown in Figure \ref{FigE6SU2}.
In the figure, we take the partial lifetime of $p \rightarrow \pi^0 + e^c$ 
as the horizontal axis and the partial lifetime of the other decay modes 
as the vertical axis.
In the Figure \ref{FigE6SU2}, we show the predictions of the $E_6$ model in the previous paper
as well as those of $E_6 \times SU(2)_F$ model.

We have two comments on these results.
First, in many model points of the $E_6\times SU(2)_F$ model the lifetime of the $p \rightarrow \pi^0 + e^c$ mode 
is shorter than the lifetimes in the $E_6$ model.
This result comes from larger $(L_e)_{11}$, because $s_{12}^{eL}$ is smaller and 
$s_{13}^{eL}$ is vanishing.
The smaller $s_{12}^{eL}$ and vanishing $s_{13}^{eL}$ are caused by
the last relation in (\ref{eq:con3}) and the last relation in (\ref{eq:con1}), respectively.
On the other hand, the lifetime of the $p \rightarrow K^0 + \mu^c$ mode does not 
become short because $(L_{e})_{22}$ does not become larger.
This is because the $s_{23}^{eL}$ has large value because of the first relation in
(\ref{eq:con3}) though $s_{12}^{eL}$ is smaller.
Second, the lifetimes for $K^0+e^c$ mode and $\pi^0+\mu^c$ mode become longer,
which can be seen in the Figure \ref{FigE6SU2}.
This is also because of smaller $s_{12}^{eL}$.

\clearpage

\subsection{Calculation of $R_1$ and $R_2$ in $E_6\times SU(2)_F\times U(1)_A$}
In the previous paper\cite{nucleon decay in U(1)A}, we have emphasized that the parameters
$R_1 = \frac{\Gamma_{n \rightarrow \pi^0 + \nu^c}}{\Gamma_{p \rightarrow \pi^0 + e^c}}$
and
$R_2 = \frac{\Gamma_{p \rightarrow K^0 + \mu^c}}{\Gamma_{p \rightarrow \pi^0 + e^c}}$
are useful to identify the grand unification groups, $SU(5)$, $SO(10)$, or $E_6$, in the
 anomalous $U(1)_A$ GUTs.
The $R_1$ can be important to know the largeness of the rank of the unification group \cite{SU5 VS SO10}.
The $R_2$ has sensitivity of the Yukawa structure, especially, for the second generation fields.

\begin{figure}[htb]
 \centering
 \includegraphics[width=12cm,clip]{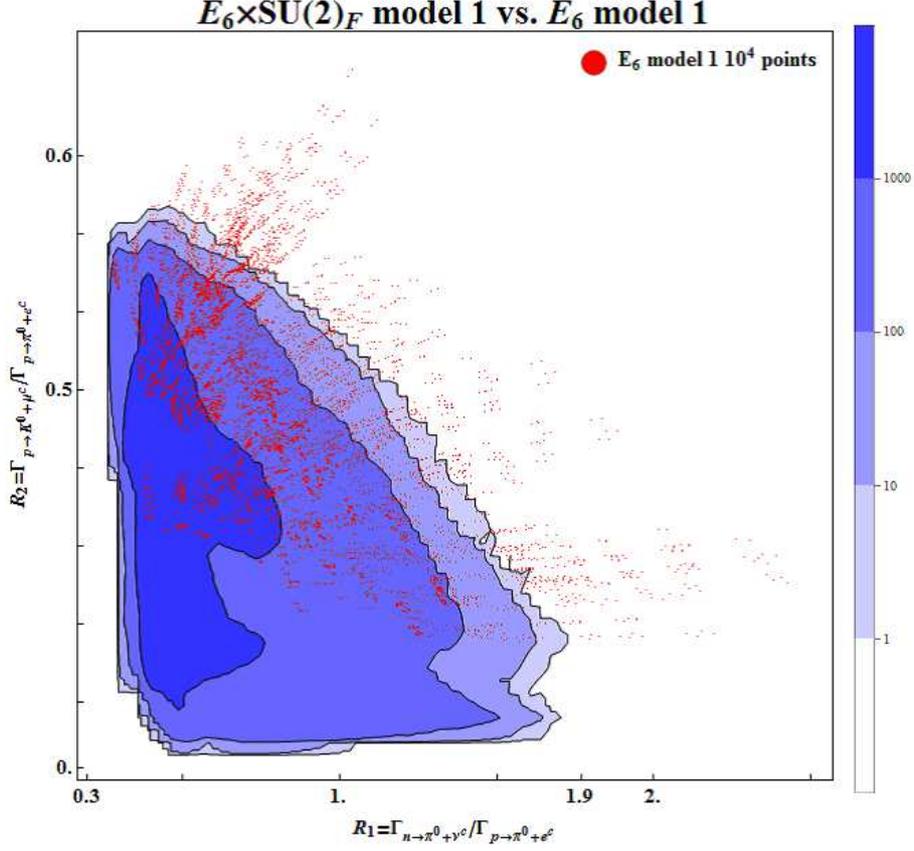}
 \caption{Contour plot
of $E_6\times SU(2)_F$ model point density. The model point density is defined by the number of
model points par unit area $(\Delta R_1, \Delta R_2)=((1.9-0.3)/50=0.032, 0.6/50=0.012)$
in $(R_1, R_2)$ plain after generating $10^6$ model points. $10^4$ model points for $E_6$ model
 without $SU(2)_F$, which are calculated in Ref. \cite{nucleon decay in U(1)A}, are dotted. 
VEVs are taken as $x=1\times 10^{16}$ GeV, $v_c=5\times 10^{14}$ GeV, and 
$v_{\phi}=5\times 10^{15}$ GeV. 
}
 \label{Fig2}
\end{figure}

\begin{figure}[hbt]
 \centering
 \includegraphics[width=12cm,clip]{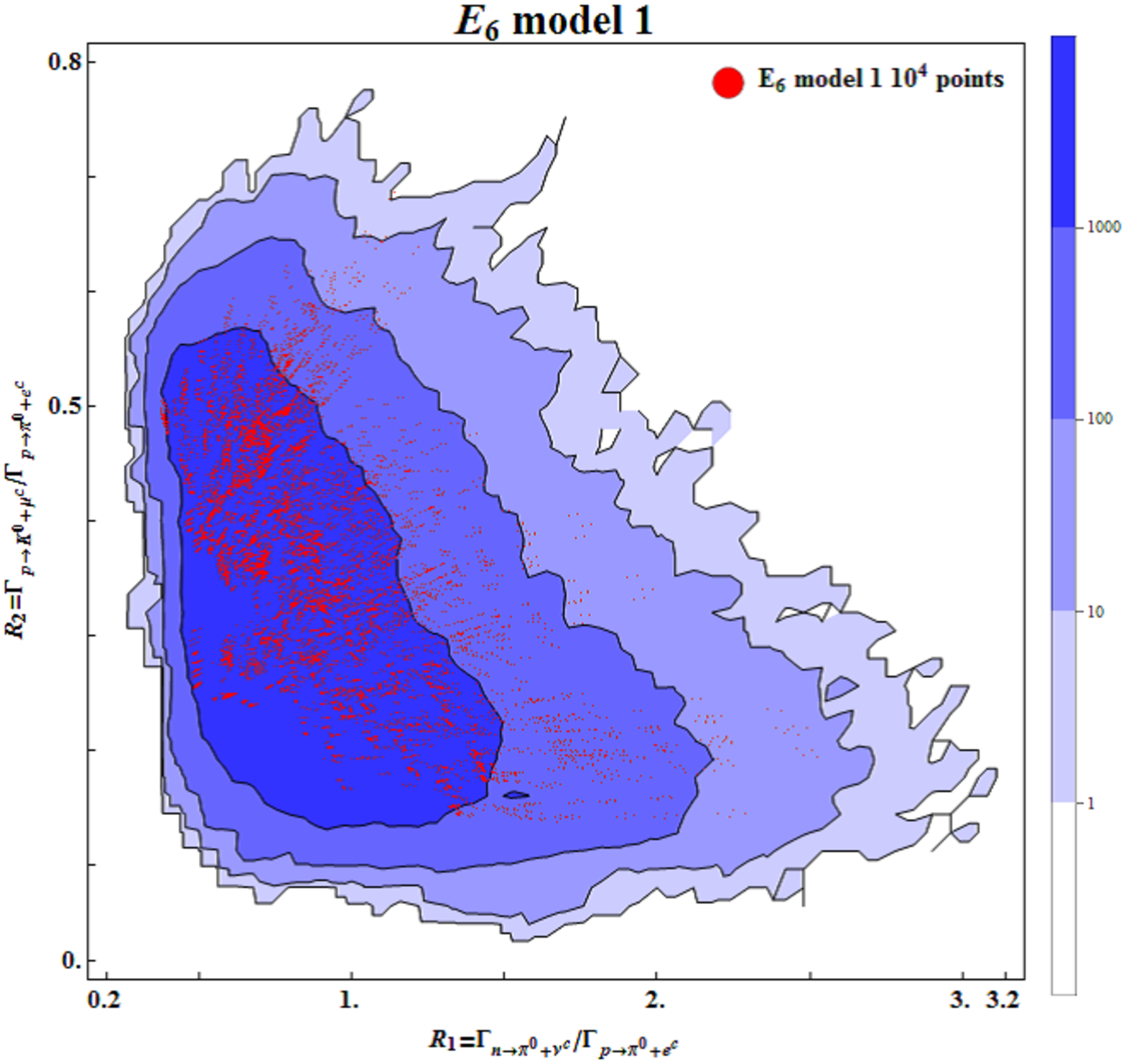}
 \caption{Contour plot 
 of model point density for the $E_6$ model without $SU(2)_F$. 
 The model point density is defined by the number of model points par unit area
$(\Delta R_1, \Delta R_2)=(0.06, 0.012)$ in $(R_1, R_2)$ plain after generating $10^6$ model points. 
 $10^4$ model points, which are  calculated in Ref. \cite{nucleon decay in U(1)A}, are dotted.
VEVs are taken as $x=1\times 10^{16}$ GeV, $v_c=5\times 10^{14}$ GeV, and 
$v_{\phi}=5\times 10^{15}$ GeV. 
}
 \label{Fig3}
\end{figure}

We have calculated these parameters for $10^6$ model points for $E_6\times SU(2)_F$ model, which
are much larger than in the previous paper\cite{nucleon decay in U(1)A}.
The results are shown in Figure \ref{Fig2}, in which the darker region represents larger density of
model points. The model points for $E_6$ model without $SU(2)_F$, which has been calculated in the
previous paper \cite{nucleon decay in U(1)A},  are dotted in the figure.
The region in which both $R_1$ and $R_2$ are small are allowed in $E_6\times SU(2)_F$ model but looks
not to be allowed in $E_6$ model without $SU(2)_F$. Of course, this can happen because the predictions of 
the two models are different. However, it is also plausible that the allowed region 
in $E_6\times SU(2)_F$ model is included in the allowed region in $E_6$ model without $SU(2)_F$ if
more model points are taken into account.
Therefore, we have re-calculated the allowed region by using 100 times more model points for 
$E_6$ model without $SU(2)_F$ (see Figure \ref{Fig3}). The allowed region for $E_6\times SU(2)_F$ model
 is almost included in the allowed region for $E_6$ model without $SU(2)_F$, though small region with
 small $R_1$ and $R_2$ is still not included. 
Since it has been found that increasing model points are important, we re-examine the conditions 
for identification of the grand unification group, which were discussed in the previous paper, 
with 100 times more model points in the next subsection.


We have two comments in Figures \ref{Fig2} and \ref{Fig3}.
First, in the $E_6 \times SU(2)_F$ model, many model points have smaller $R_1$ and $R_2$ 
than these in the $E_6$ model without $SU(2)_F$.
This is because $\Gamma_{p \rightarrow \pi^0 + e^c}$ is tend to be larger due to  the small mixings 
between the electron and the other charged leptons as we mentioned in the previous subsection.
Second, in the $E_6 \times SU(2)_F$ model, the plotted region in the $R_1$ and $R_2$ plain becomes
smaller than in the $E_6$ model, as expected.
This is because the diagonalizing matrices are restricted by 
relations (\ref{eq:con1})-(\ref{eq:con4}).

\clearpage

\subsection{Identification of GUT models} 
In this subsection we re-examine the conditions for identification of the grand unification group
by using $10^6$ model points which are 100 times more than in the previous 
paper \cite{nucleon decay in U(1)A}.

\begin{figure}[hbt]
 \centering
 \includegraphics[width=12cm,clip]{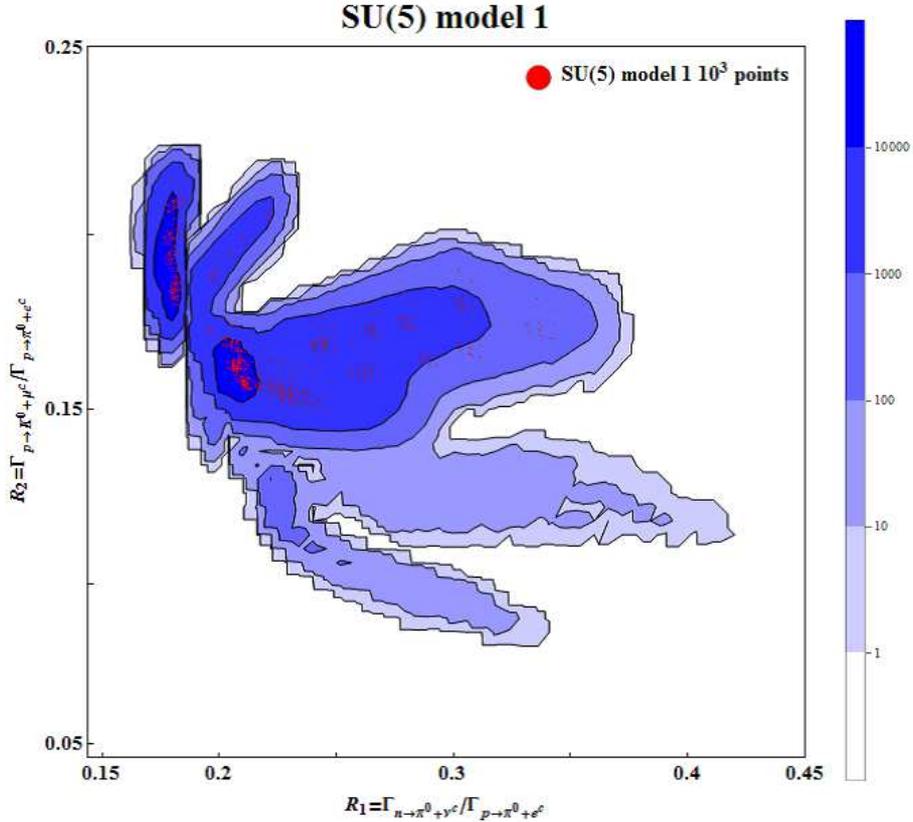}
 \caption{Contour plot of $SU(5)$ model point density. The model point density is defined
 by number of model points par unit area $(\Delta R_1, \Delta R_2)=(0.006, 0.004)$
 in $(R_1, R_2)$ plain after generating $10^6$ model points. 
 $10^4$ model points, which are  calculated in Ref. \cite{nucleon decay in U(1)A}, are dotted.
VEVs are taken as $x=1\times 10^{16}$ GeV. 
}
 \label{Fig4}
\end{figure}

In order to examine the statement that the unification group is not $SU(5)$ if $R_1>0.4$,
we have calculated $R_1$ and $R_2$ in $SU(5)$ model with $10^6$ model points (see Figure \ref{Fig4}).
The Figure shows that there are very few model points with $R_1>0.4$. Therefore, the statement is
almost satisfied even if $10^6$ model points are taken into account.

\begin{figure}[hbt]
 \centering
 \includegraphics[width=12cm,clip]{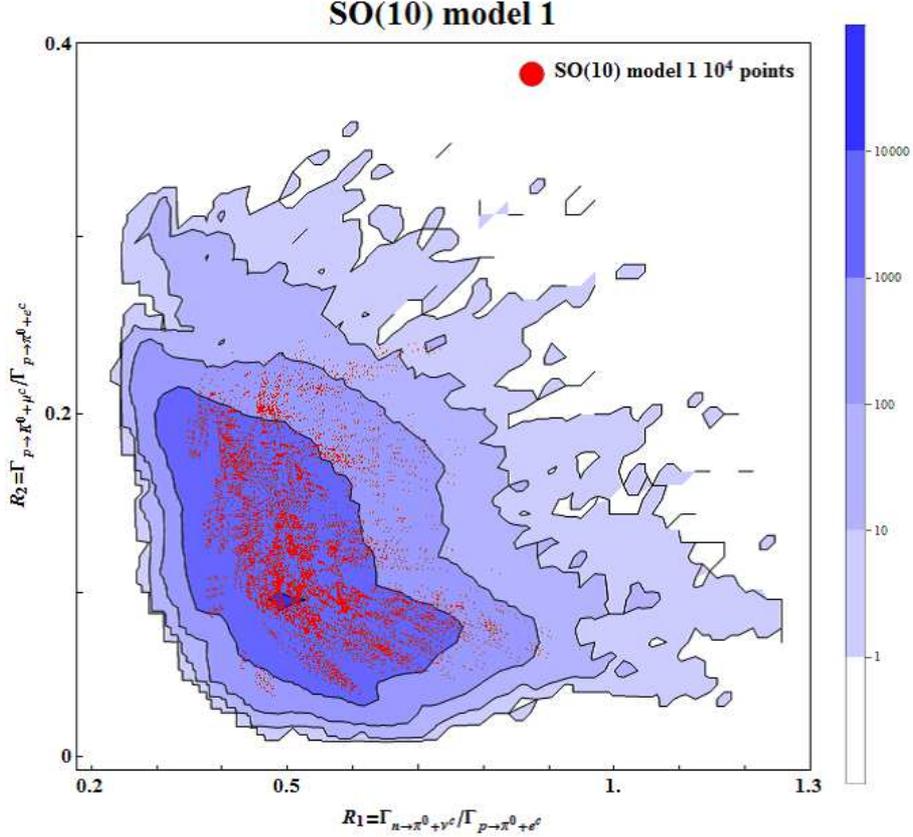}
 \caption{Contour plot of $SO(10)$ model point density. The model point density is defined
 by number of model points par unit area $(\Delta R_1, \Delta R_2)=(0.022, 0.008)$
 in $(R_1, R_2)$ plain after generating $10^6$ model points.
 $10^4$ model points, which are  calculated in Ref. \cite{nucleon decay in U(1)A}, are dotted.
VEVs are taken as $x=1\times 10^{16}$ GeV, $v_c=5\times 10^{14}$ GeV. 
}
 \label{Fig5}
\end{figure}

In order to examine the statements that the unification group is $E_6$ if $R_1>1$ and that the
unification group is implied to be $E_6$ if $R_2>0.3$, we have calculated $R_1$ and $R_2$ in
$SO(10)$ model with $10^6$ model points (see Figure \ref{Fig5}). Note that
the effect of $SO(10)$ X-type gauge boson $X'$ becomes almost maximal with the VEVs adopted in the calculation.
The Figure shows that there are very few model points with $R_1>1$ or with $R_2>0.3$. 
Therefore, these statements are
almost satisfied even if $10^6$ model points are taken into account.



\begin{figure}[htb]
 \centering
 \includegraphics[width=12cm,clip]{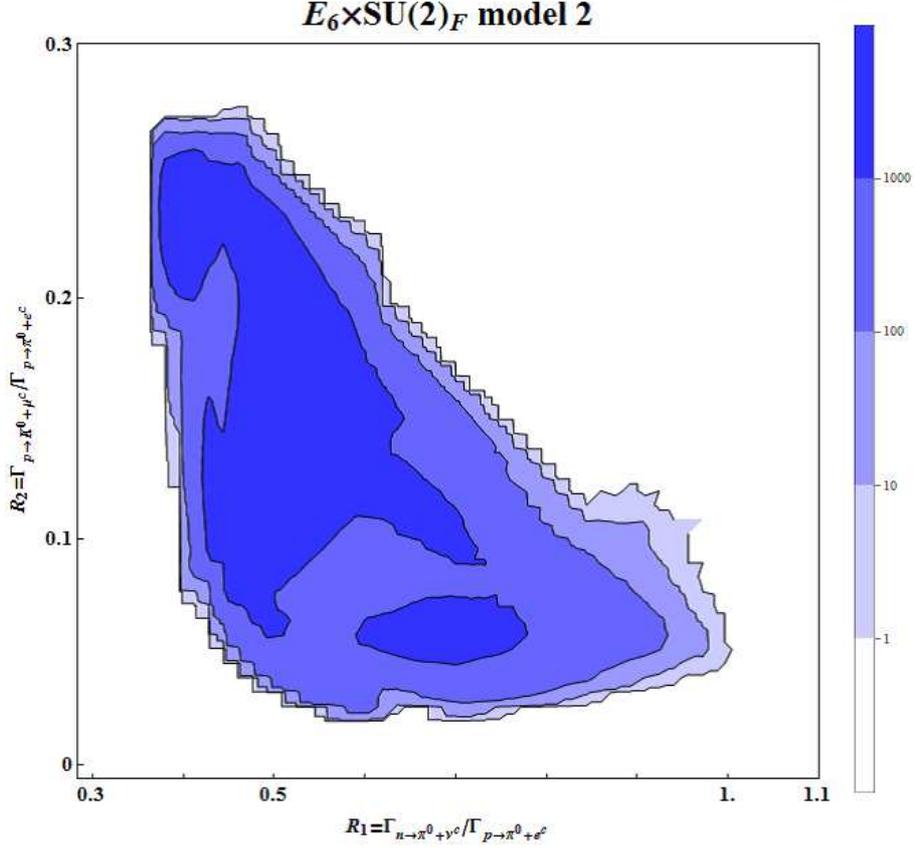}
 \caption{Contour plot of the model point density of $E_6\times SU(2)_F$ model 2. The model point density
 is defined by the number of model points par unit area  $(\Delta R_1, \Delta R_2)=(0.016, 0.006)$
 in $(R_1, R_2)$ plain after generating $10^6$ model points.
VEVs are taken as $x=5\times 10^{15}$ GeV, $v_c=5\times 10^{14}$ GeV, and 
$v_{\phi}=5\times 10^{15}$ GeV. 
}
 \label{Fig6}
\end{figure}

In the end of this subsection, we show the result in $E_6 \times SU(2)_F$ model with $x=5\times 10^{15}$
GeV.
The difference is only the VEV of adjoint Higgs.
As seen in Figure \ref{Fig6}, the  $E_6 \times SU(2)_F$ with smaller $x$ predicts smaller $R_1$ and $R_2$
than the original $E_6 \times SU(2)_F$ model which has $x=1\times 10^{16}$ GeV.
This is because the nucleon decay via dimension-6 operators which is induced by $X''$ 
exchange is more suppressed in $E_6 \times SU(2)_F$ model with smaller $x$ than in the model with larger $x$.

\clearpage

\section{Discussion and Summary}
In this paper we have calculated the partial lifetime for various decay modes of the  nucleons
via dimension-6 operators 
in anomalous $U(1)_A$ $E_6 \times SU(2)_F$ SUSY GUT model with the spontaneously broken CP symmetry.
Once we fix the VEVs of GUT Higgs, the main ambiguities come from the diagonalizing matrices of
quark and lepton mass matrices. Since the $SU(2)_F$ symmetry can reduce the ambiguities, the predictions
have become more restricted than the $E_6$ model without $SU(2)_F$ symmetry.
We have derived the various relations on the components of the diagonalizing matrices
from the constraints on the Yukawa couplings which are realized in $E_6\times SU(2)_F$ model.
Among the relations, we have used 9 relations which are not dependent on the renormalization scale.
We have showed that only 6 parameters are sufficient to fix the 7 diagonalizing $3\times 3$ matrices. 

In this calculation, we have increased the model points up to $10^6$ from $10^4$ in the previous paper.
Even with such many model points, the previous conclusion is still valid, that  
$R_1 = \frac{\Gamma_{n \rightarrow \pi^0 + \nu^c}}{\Gamma_{p \rightarrow \pi^0 + e^c}}$
and
$R_2 = \frac{\Gamma_{p \rightarrow K^0 + \mu^c}}{\Gamma_{p \rightarrow \pi^0 + e^c}}$
are useful to identify the grand unification groups, $SU(5)$, $SO(10)$, or $E_6$, in the 
anomalous $U(1)_A$ GUTs.

It is important to consider how to test the GUT models. The most important prediction of the 
GUT is the nucleon decay, and therefore, the calculations for the partial decay widths for
various GUT models are important. One more interesting evidence of the GUT models may appear in
the SUSY breaking parameters, especially, in scalar fermion masses through the $D$-term
contribution which are generated if the rank of the unification group is larger than 4 or
the additional gauge symmetry like $SU(2)_F$ is introduced \cite{D-term}. 
We will study this possibility in the $E_6\times SU(2)_F$ models in future.
The estimation of the diagonalizing matrices in this paper must be important in predicting 
the FCNC processes induced by the non-vanishing D-term.

\section{Acknowledgement}
N.M. is supported in part by Grants-in-Aid for Scientific Research from MEXT of 
Japan. This work is partially supported by the Grand-in-Aid for Nagoya University
Leadership Development Program for Space Exploration and Research Program from the MEXT 
of Japan.

\appendix

\section{The diagonalizing of Yukawa matrices (in leading order)}
\label{sec:diagonalizing}
Hereafter, we summarize how to diagonalize the $3 \times 3$ matrix $Y_{ij}$.
In this calculation we suppose that the Yukawa matrix has hierarchies, 
$Y_{ij} \ll Y_{kj}$ and $Y_{ij} \ll Y_{il}$ when $i < k$ and $j < l$.
References for this calculation is \cite{E6SU2_2,diagonalizing}.

Diagonalizing the Yukawa matrix, we translate the flavor eigenstate  $\psi$ 
into mass eigenstate $\psi'$.
We make the Yukawa matrix $Y$ diagonal, as
\begin{align}
\psi_{L i} Y_{ij} \psi_{Rj}^c &= 
(L_{\psi}^{\dagger} \psi_{L})_i (L_{\psi}^T Y R_{\psi})_{ij} (R_{\psi}^{\dagger}\psi_{R}^c)_j \\
 &=\psi'_{L i} Y_{diag\: ij} \psi'^c_{Rj} \nonumber
\label{eq:diagonalizing}
\end{align}
where unitary matrices $L_{\psi}$ and $R_{\psi}$ are the diagonalizing matrices, 
and $i$, $j$ ($i,\: j=1,\:2,\:3$) are the index of generation.

We express parameters for diagonalizing matrix $L$ and $R$ as
\begin{align}
 L^T &\equiv
  \begin{pmatrix}
   c_{12}^L & -s_{12}^L & 0 \\
   s_{12}^{L*} & c_{12}^L & 0 \\
   0 & 0 & 1
  \end{pmatrix}
  \begin{pmatrix}
   c_{13}^L & 0 & -s_{13}^L \\
   0 & 1 & 0 \\
   s_{13}^{L*} & 0 & c_{13}^L
  \end{pmatrix}
  \begin{pmatrix}
   1 & 0 & 0 \\
   0 & c_{23}^L & -s_{23}^L \\
   0 & s_{23}^{L*} & c_{23}^L
  \end{pmatrix}
 \equiv P_{12}^L P_{13}^L P_{23}^L, \\
 R &\equiv
 \begin{pmatrix}
  1 & 0 & 0 \\
  0 & c_{23}^R & s_{23}^R \\
  0 & -s_{23}^{R*} & c_{23}^R
 \end{pmatrix}
 \begin{pmatrix}
  c_{13}^R & 0 & s_{13}^R \\
  0 & 1 & 0 \\
  -s_{13}^{R*} & 0 & c_{13}^R
 \end{pmatrix}
 \begin{pmatrix}
  c_{12}^R & s_{12}^R & 0 \\
  -s_{12}^{R*} & c_{12}^R & 0 \\
  0 & 0 & 1
 \end{pmatrix}
 \equiv P_{23}^{R\dagger} P_{13}^{R\dagger} P_{12}^{R\dagger}
\end{align}
where $s_{ij}^{L/R} \equiv \sin\theta_{ij}^{L/R}e^{i\chi_{ij}^{L/R}}$ and 
$c_{ij}^{L/R} \equiv \cos\theta_{ij}^{L/R}$.
We define the Yukawa matrix $Y$ as
\begin{equation}
 Y \equiv
  \begin{pmatrix}
   y_{11} & y_{12} & y_{13} \\
   y_{21} & y_{22} & y_{23} \\
   y_{31} & y_{32} & y_{33}
  \end{pmatrix}.
\end{equation}
The Yukawa matrix is diagonalized as
\begin{align}
 L^T Y R &= P_{12}^LP_{13}^LP_{23}^L
 \begin{pmatrix}
  y_{11} & y_{12} & y_{13} \\
  y_{21} & y_{22} & y_{23} \\
  y_{31} & y_{32} & y_{33}
 \end{pmatrix}
 P_{23}^{R\dagger}P_{13}^{R\dagger}P_{12}^{R\dagger} \nonumber \\
 &\simeq P_{12}^LP_{13}^L
 \begin{pmatrix}
  y_{11} & y^{\prime}_{12} & y_{13} \\
  y^{\prime}_{21} & y^{\prime}_{22} & 0 \\
  y_{31} & 0 & y_{33}
 \end{pmatrix}
 P_{13}^{R\dagger}P_{12}^{R\dagger} \notag \\
 &\simeq P_{12}^L
 \begin{pmatrix}
  y^{\prime}_{11} & y^{\prime}_{12} & 0 \\
  y^{\prime}_{21} & y^{\prime}_{22} & 0 \\
  0 & 0 & y_{33}
 \end{pmatrix}
 P_{12}^{R\dagger} =
 \begin{pmatrix}
  y^{\prime\prime}_{11} & 0 & 0 \\
  0 & y^{\prime}_{22} & 0 \\
  0 & 0 & y_{33}
 \end{pmatrix}.
\end{align}
In the calculation, we use the approximation that the mixing angles are small, i.e.,  
$|s_{ij}^{L/R}| \sim |\theta_{ij}| \ll 1$ $(s_{ij}^{L/R} \sim \theta_{ij} 
e^{i\chi_{ij}^{L/R}})$ and
$c_{ij}^{L/R} \simeq 1$.
Mixing angles of the diagonalizing matrix and eigenvalues are estimated in this assumption as
\begin{equation}
 y^{\prime}_{22} \simeq y_{22} - \frac{y_{23}y_{32}}{y_{33}},\qquad
  y^{\prime}_{12} \simeq y_{12} - \frac{y_{13}y_{32}}{y_{33}},\qquad
  y^{\prime}_{21} \simeq y_{21} - \frac{y_{23}y_{31}}{y_{33}},
\end{equation}
\begin{equation}
 y^{\prime}_{11} \simeq y_{11} - \frac{y_{13}y_{31}}{y_{33}},\qquad
  y^{\prime\prime}_{11} \simeq y_{11}^{\prime}
  - \frac{y_{12}^{\prime}y_{21}^{\prime}}{y_{22}^{\prime}}.
\end{equation}
\begin{equation}
 s_{23}^L \simeq \frac{y_{23}}{y_{33}},\qquad
  s_{13}^L \simeq \frac{y_{13}}{y_{33}},\qquad
  s_{12}^L \simeq
  \frac{y_{12}y_{33} - y_{13}y_{32}}{y_{22}y_{33} - y_{23}y_{32}}.
\end{equation}
\begin{equation}
 s_{23}^{R*} \simeq \frac{y_{32}}{y_{33}},\qquad
  s_{13}^{R*} \simeq \frac{y_{31}}{y_{33}},\qquad
  s_{12}^{R*} \simeq
  \frac{y_{21}y_{33} - y_{31}y_{23}}{y_{22}y_{33} - y_{23}y_{32}}.
\end{equation}

For reference, we explain the diagonalization for $2 \times 2$ matrix without approximation.
\begin{equation}
 \begin{pmatrix}
  c_L & -s_L \\
  s_L^* & c_L
 \end{pmatrix}
 \begin{pmatrix}
  y_{11} & y_{12} \\
  y_{21} & y_{22}
 \end{pmatrix}
 \begin{pmatrix}
  c_R & s_R \\
  -s_R^* & c_R
 \end{pmatrix}
 =
 \begin{pmatrix}
  y^{\prime}_{11} & 0 \\
  0 & y^{\prime}_{22}
 \end{pmatrix}.
\end{equation}
As we defined above, we define that $s_{L/R} \equiv \sin\theta_{L/R}e^{i\chi_{L/R}}$ 
and $c_{L/R} \equiv \cos\theta_{L/R}$.
$\tan 2\theta_{L/R}$ are taken as
\begin{align}
 \tan 2\theta_L &=
 \frac{2(y_{12}y_{22} + y_{11}y_{21}e^{2i\chi_R})}
 {y_{22}^2e^{i\chi_L} - y_{11}^2e^{-i(\chi_L - 2\chi_R)}
 + y_{21}^2e^{i(\chi_L + 2\chi_R)} - y_{12}^2e^{-i\chi_L}}, \\
 \tan 2\theta_R &=
 \frac{2(y_{11}y_{12} + y_{21}y_{22}e^{2i\chi_L})}
 {y_{22}^2e^{-i(\chi_R - 2\chi_L)} - y_{11}^2e^{i\chi_R}
 - y_{21}^2e^{i(\chi_R + 2\chi_L)} + y_{12}^2e^{-i\chi_R}}.
\end{align}
And eigenvalues become
\begin{align}
 y^{\prime}_{11} &= y_{12} c_L (\frac{y_{11}}{y_{12}}c_R -s_R^*)
 -y_{22} s_L(\frac{y_{21}}{y_{22}}c_R -s_R^*), \\
 y^{\prime}_{22} &= y_{12} s_L^* (\frac{y_{11}}{y_{12}}s_R +c_R)
 +y_{22} c_L(\frac{y_{21}}{y_{22}}s_R +c_R).
\end{align}
When the $2 \times 2$ matrix has hierarchy, 
$y_{11} \ll y_{12} \sim y_{21} \ll y_{22}$, the angles and the eigenvalues are approximately obtained as
\begin{equation}
s_L^* \sim \frac{y_{12}}{y_{22}}e^{-i\chi_L},\: 
s_R \sim \frac{y_{21}}{y_{22}}e^{i\chi_R},
\end{equation}
\begin{equation}
y^{\prime}_{11} \sim y_{11}+\frac{y_{12}y_{21}}{y_{22}},\: 
y^{\prime}_{22} \sim y_{22}.
\end{equation}

The diagonalizing matrices of left-handed up-type quark and 
down-type quark, $L_u$ and $L_d$, are given by
\begin{equation}
 \label{eq:VL}
L_{u/d}^T = P_{12}^{u/dL}P_{13}^{u/dL}P_{23}^{u/dL} \simeq
  \begin{pmatrix}
   1 & -s_{12}^{u/dL} & -s_{13}^{u/dL} + s_{23}^{u/dL}s_{12}^{u/dL} \\
   s_{12}^{u/dL*} & 1 & -s_{23}^{u/dL} \\
   s_{13}^{u/dL*} & s_{23}^{u/dL*} & 1
  \end{pmatrix}.
\end{equation}
Therefore, the CKM matrix $U_{CKM}$ is calculated as 
\begin{align}
 U_{CKM} &\equiv L_{u}^{\dagger}L_{d} \nonumber \\
   &\simeq \left(
\begin{array}{cc}
1 & s_3^{dL} - s_3^{uL}  \\
s_3^{uL*} - s_3^{dL*} & 1 \\
s_2^{uL*} - s_2^{dL*} - s_3^{dL*}(s_1^{uL*} - s_1^{dL*}) & s_3^{uL*} - s_3^{dL*}
\end{array}\right. \nonumber \\
& \hspace{6cm} \left.
\begin{array}{c}
s_2^{dL} - s_2^{uL} - s_3^{uL}(s_1^{dL} - s_1^{uL}) \\
s_1^{dL} - s_1^{uL} \\
1
\end{array}\right) \nonumber \\
  &\equiv 
  \begin{pmatrix}
   1 & U_{us} & U_{ub} \\
   -U_{us}^* & 1 & U_{cb} \\
   U_{us}^*U_{cb}^* - U_{ub}^* & -U_{cb}^* & 1
  \end{pmatrix}.
\end{align}
We can also calculate the MNS matrix by replacement, 
$u\leftrightarrow \nu$ and $d\leftrightarrow e$.

\end{document}